\documentclass[a4paper]{revtex4}
\pdfoutput=1
\usepackage{amsmath, amsthm, amsfonts, amssymb, latexsym}
\usepackage{graphicx}
\usepackage{fancyhdr}

\usepackage{hyperref}
\usepackage{amsmath}
\usepackage{cancel}
\usepackage{color}

\begin{document}

\title{Two-loop stability of singlet extensions of the SM with dark matter}

%

\author{Raul Costa}
\affiliation{Centro de F\'\i sica Te\'orica e Computacional,
Universidade de Lisboa 
1649-003 Lisboa, Portugal} 
\author{Ant\'onio P. Morais} 
\affiliation{Departamento de F\'\i sica da Universidade de Aveiro and CIDMA, 
Campus de Santiago, 3810-183 Aveiro, Portugal}
\affiliation{Centro de F\'\i sica Te\'orica e Computacional,
Universidade de Lisboa
1649-003 Lisboa, Portugal} 
\author{Marco O. P. Sampaio}
\affiliation{Departamento de F\'\i sica da Universidade de Aveiro and CIDMA \\ 
Campus de Santiago, 3810-183 Aveiro, Portugal}
\author{Rui Santos}
\affiliation{Centro de F\'\i sica Te\'orica e Computacional,
Universidade de Lisboa
1649-003 Lisboa, Portugal} 
\affiliation{
 Instituto Superior de Engenharia de Lisboa - ISEL,
 1959-007 Lisboa, Portugal
} 

\begin{abstract}
We present a study of the high energy stability of a minimal complex singlet extension of the Standard Model with or without dark matter (CxSM). We start by obtaining the beta functions of the couplings of the theory from the effective potential and then perform the RGE evolution for the allowed parameter space of the model at the electroweak scale. We obtain the scale up to which the model survives and combine this information with all the LHC measurements as well as bounds from dark matter detection experiments as well as the relic density best measurement from Planck. We conclude that scenarios which solve both the dark matter and stability problems must contain a dark particle heavier than 50~GeV and a new visible state (mixing with the SM-like Higgs) heavier than 170~GeV.
\end{abstract}

\maketitle

\thispagestyle{fancy}


\section{Introduction}
The recent success of the Higgs boson discovery at the CERN Large Hadron Collider (LHC) experiments~\cite{ATLAS:2012ae,Chatrchyan:2012tx}, while putting the Standard Model (SM) on a firmer ground, is pushing us towards finding solutions to outstanding problems which are not yet addressed in the model. Such are the question of the nature of dark matter, what is the mechanism generating the observed matter anti-matter asymmetry or even what solves the apparent metastability of the SM Higgs potential~\cite{Degrassi:2012ry}. 

A partial answer to these questions may be about to appear in the next runs of the LHC, and could manifest itself in a minimal form. An example of a minimal parametrisation which naturally connects to the Higgs sector under scrutiny at the LHC is given by the scalar singlets framework. This provide, in addition, quite a natural connection into hidden sectors~\cite{Patt:2006fw}. In fact, in the SM, there are only three gauge singlets that we can build with dimension less than four. In particular, the only scalar is $H^\dagger H$ ($H$ denotes the Higgs doublet), which means that scalar singlet fields naturally couple to the Higgs through this operator
\begin{equation}
V=V_{\rm SM}(H^\dagger H)+H^\dagger H \times\mathcal{O}^{(2)}_\delta(\phi_i)+V_{\rm new}(\phi_i) \; .
\end{equation}
Here we split the potential in three parts: i) the SM-like part, $V_{\rm SM}$; ii) The SM Higgs singlet operator coupling to some operator, $\mathcal{O}^{(2)}_\delta$,  with mass dimension up to 2, which depends on the new singlet fields $\phi_{i}$, and iii) the new (purely) singlet potential $V_{\rm new}$. In this class of models, the physical mass states couple to the SM only through the Higgs field fluctuation, if they mix, otherwise they are dark particles. Their mixing factors  ($\kappa_a$, $a=1,\ldots,\#{\rm scalars}$) suppress their couplings to other SM particles compared to a SM like Higgs coupling. At tree level, if we denote the SM Higgs boson fluctuation by $h$ and the mass eigenstates by $H_a$, then we have a sum rule for the $\kappa_a$~\cite{Barger:2009me}
\begin{equation}
h=\sum_a\kappa_aH_a\;\;, \;\; \sum_a|\kappa_a|^2=1\; .
\end{equation}
In this study we focus on complex singlet (i.e. two scalar) singlet extension which provides  both a simple parametrisation of dark matter (through the state that do not mix with the SM Higgs fluctuation) and can be made compatible with electroweak baryogenesis as to generate the matter anti-matter asymmetry~\cite{Menon:2004wv, Huber:2006wf, Profumo:2007wc,Barger:2011vm, Espinosa:2011ax,Coimbra:2013qq,Costa:2014qga}. We will find that, in addition, such model can also improve the stability of the SM at high energies~\cite{Costa:2014qga}.

\section{A minimal complex singlet model with dark matter}
We consider a model where the SM is extended by a complex scalar singlet $\mathbb{S}=(S+i A)/\sqrt{2}$ with potential~\cite{Coimbra:2013qq}
\begin{equation}
V=\frac{m^2}{2}H^\dagger H+\frac{\lambda}{4}(H^\dagger H)^2+\frac{\delta_2}{2}H^\dagger H |\mathbb{S}|^2+\frac{b_2}{2}|\mathbb{S}|^2+\frac{d_2}{4}|\mathbb{S}|^4+\left(\frac{b_1}{4}\mathbb{S}^2+a_1\mathbb{S}+c.c.\right) \; .
\end{equation}
This can be seen as a softly broken version of a $U(1)$ symmetric model which preserve a $\mathbb{Z}_2$ parity for the imaginary part, $A$. This minimal complex singlet model contains two possible phases which break electroweak symmetry as necessary in the SM. If we define the vacuum expectation values, $v_i$, of the new fields through $S\equiv v_S+s$, $A\equiv v_A+a$, while keeping the SM Higgs vev $v$ then we have:
\begin{itemize}
\item $\mathbb{Z}_2$ phase ($v_S\neq 0,v_A=0$) -- symmetric phase: Here the two fields $h,S$ mix to form the two visible mass eigenstates and $A$ is a dark matter candidate. We highlight $\kappa_{126}=\cos\alpha$ and $\kappa_{\rm new}=\sin\alpha$ in the mixing matrix
\begin{equation}
\left(\begin{array}{c}
H_{126} \\
H_{\rm new} \\ 
H_{\rm DM} \end{array}\right)=
\left(\begin{array}{ccc} \kappa_{126} &-\sin\alpha &0 \\ 
\kappa_{\rm new} &\cos\alpha &0 \\ 
0& 0&1 \end{array}\right)
\left(\begin{array}{c}h \\ s \\ a\end{array}\right) \; .
\end{equation}
\item $\cancel{\mathbb{Z}_2}$ phase ($v_S\neq 0,v_A\neq0$) -- spontaneously broken phase: All scalar massive degrees of freedom are visible and mix. Again we highlight the $\kappa_i$ in the mixing elements $R_{ij}$
\begin{equation}
\left(\begin{array}{c} H_{126} \\ 
H_{\rm light} \\ 
H_{\rm Heavy} \end{array}\right)
=\left(\begin{array}{ccc}\kappa_{126} &R_{1s} & R_{1a}  \\ 
\kappa_{\rm light} &R_{2s} & R_{2a} \\ 
\kappa_{\rm Heavy} &R_{3s} & R_{3a} 
\end{array}\right)\left(\begin{array}{c}h \\ s \\ a \end{array}\right) \; .
\end{equation}
\end{itemize}
 In both phases we match the first visible scalar to have the mass of the recently found Higgs. In the broken phase, the other two are ordered in mass relative to each other (but they can both be lighter or heavier than the Higgs or one lighter one heavier than the Higgs). This minimal model is representative of all kinematically interesting situations in terms of visible and invisible channels in a sector with two new real singlet degrees of freedom at the LHC.

Finally it is important to note that, in this model, observables such as for example cross-sections, can be easily related to SM observables through the $\kappa_i$ factors (for example the production cross section for one of these scalars is $\sigma(H_i)=\kappa_i^2\sigma_{\rm SM}(H_i)$ -- see also~\cite{Costa:2014qga}).

\subsection{Renormalisation group equations and stability}

In this study, our perspective was to perform a global scan over all possible scenarios combining the phenomenological constraints, at the electroweak scale, with the requirement that the theory remains stable up to a high energy scale (such as the GUT or Planck scale). Previous global scans for this model where mostly tree level~\cite{Coimbra:2013qq,Costa:2014qga}, and one-loop studies focused on particular points/scenarios~\cite{Gonderinger:2009jp}. 

To obtain the evolution equations for the couplings of the theory, to run their input values from the electroweak scale to high energies, one needs to compute loop corrections of the following two quantities (to keep track of the loop order, we define $\varepsilon\equiv \hbar/(16\pi^2)$):
\begin{equation}
\begin{cases}V_{\rm eff}=V^{(0)}+ \varepsilon V^{(1)}+\varepsilon^2 V^{(2)}+\ldots \vspace{3mm}\\
G^{-1}_{ij}=\Pi^{(0)}_{ij}+\varepsilon \Pi_{ij}^{(1)}+\varepsilon^2 \Pi_{ij}^{(2)}+\ldots
\end{cases} \hspace{-6mm}\xrightarrow[divs.]{\frac{dV_{\rm eff}}{dt}=0} 
\begin{cases}\frac{dL}{dt}=\varepsilon \beta^{(1)}_{L}+\varepsilon^2 \beta^{(2)}_{L}+\ldots \vspace{3mm} \\ \frac{1}{v_i}\frac{d v_i}{dt}= \varepsilon \gamma^{(1)}_{i}+\varepsilon^2 \gamma^{(2)}_{i}+\ldots
 \end{cases}\label{eq:scale_invariance}
\end{equation}
In the remainder $t\equiv \log \mu$, where $\mu$ is the renormalisation energy scale. On the first line we indicate the loop expansion of the effective potential, $V_{\rm eff}$, with $V^{(n)}$ the $n$-loop order correction. The scale invariance of the effective potential provides the beta functions which determine the evolution equation of each coupling (all couplings denoted collectively by $L$ here). Using the general formalism in~\cite{Martin:2001vx} we have computed them at two loops.

The second quantity to compute (second line of Eq.~\eqref{eq:scale_invariance}) is the inverse propagator (or two point function) which encodes the anomalous dimensions ($\gamma_i^{(n)}$). These are responsible for the evolution of vacuum expectation values of fields (right hand side). Again, this is given by the tree level inverse propagator plus loop corrections, the self energies and the anomalous dimensions are extracted from their divergences~\cite{Machacek:1983tz}.

In summary we obtained the Renormalisation Group Equations (RGEs) at two loops including all SM contributions~\cite{Costa:2014qga}.

To provide low energy input for the RGE running let us consider the loop expansion of couplings (given physical input such as masses and VEVs) and corresponding beta functions at the $Z$-boson mass scale:
\begin{eqnarray}
L(M_z)&=&L^{(0)}+\varepsilon L^{(1)}(m^2_i,v_i)+\ldots  \\
\beta_L&=&\varepsilon \beta^{(1)}_{L}+\varepsilon^2\beta^{(2)}_{L}+\ldots
\end{eqnarray}
It is clear from these expressions, that if we truncate the coupling expansion at tree level (one-loop) the beta function is of one-loop (two-loop) order. In Sect.~\ref{sec:discussion} we evolve the couplings to high energies using one and two loop RGEs which we have checked with tree level and one-loop accurate low scale input data~\cite{Costa:2014qga}. In such study, in addition to imposing a correct electroweak symmetry breaking pattern and stability at the low scale, we also impose that no runaway directions are developed and that perturbative unitarity holds, at all scales. This second condition, in practice, also prevents Landau poles. Note that strict stability up to the high scale, which is what we are requiring, is not possible in the SM as shown in~\cite{Degrassi:2012ry}.

\subsection{Phenomenological constraints}

In all scans we have used the \texttt{ScannerS} code~\cite{ScannerS}, where a model class was implemented to impose all theoretical constraints (vacuum stability, boundedness from below and perturbative unitarity). In the phenomenological scans we have implemented constraints from electroweak precision variables (STU) directly in \texttt{ScannerS}, and collider constraints were applied using \texttt{HiggsBounds} (to set 95\% C.L. on new unobserved scalars) and \texttt{HiggsSignals} to obtain the probability of the model point to fit the data~\cite{Bechtle:2013wla,Bechtle:2013xfa}. All SM-like decay widths were computed with \texttt{Hdecay}~\cite{Djouadi:1997yw}. Dark matter constraints from the Planck data~\cite{Ade:2013zuv} for the relic density and limits on direct detection cross section from LUX~\cite{Akerib:2013tjd} were applied by computing the corresponding quantities for this model with \texttt{micrOMEGAS}~\cite{Belanger:2014hqa} (see also~\cite{Costa:2014qga} for details).

\section{Discussion of the results}\label{sec:discussion}
\begin{figure}
\begin{center}
\includegraphics[width=0.35\linewidth,clip=true,trim=40 35 25 45]{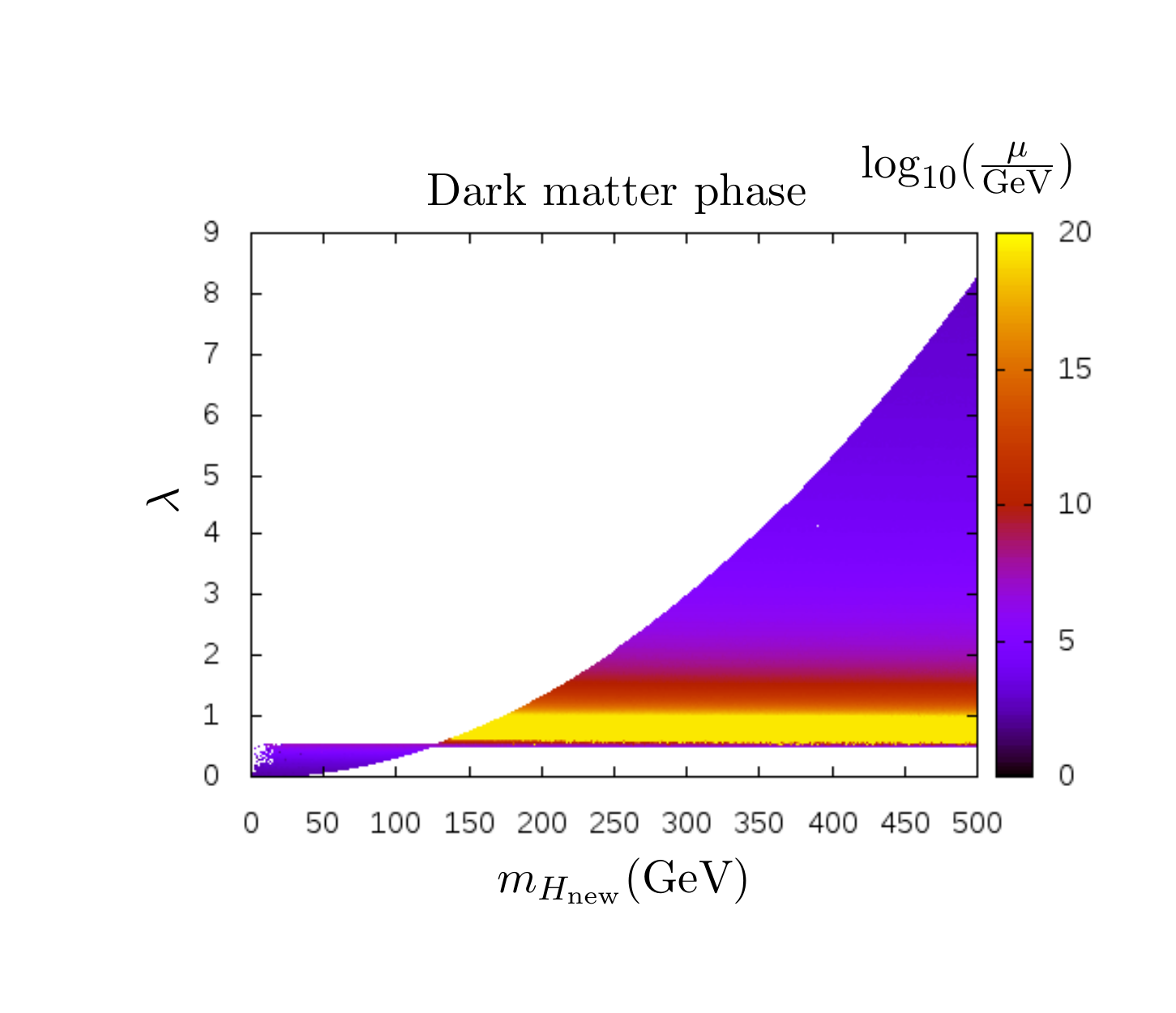}\hspace{0.1\linewidth}\includegraphics[width=0.35\linewidth,clip=true,trim=40 35 25 45]{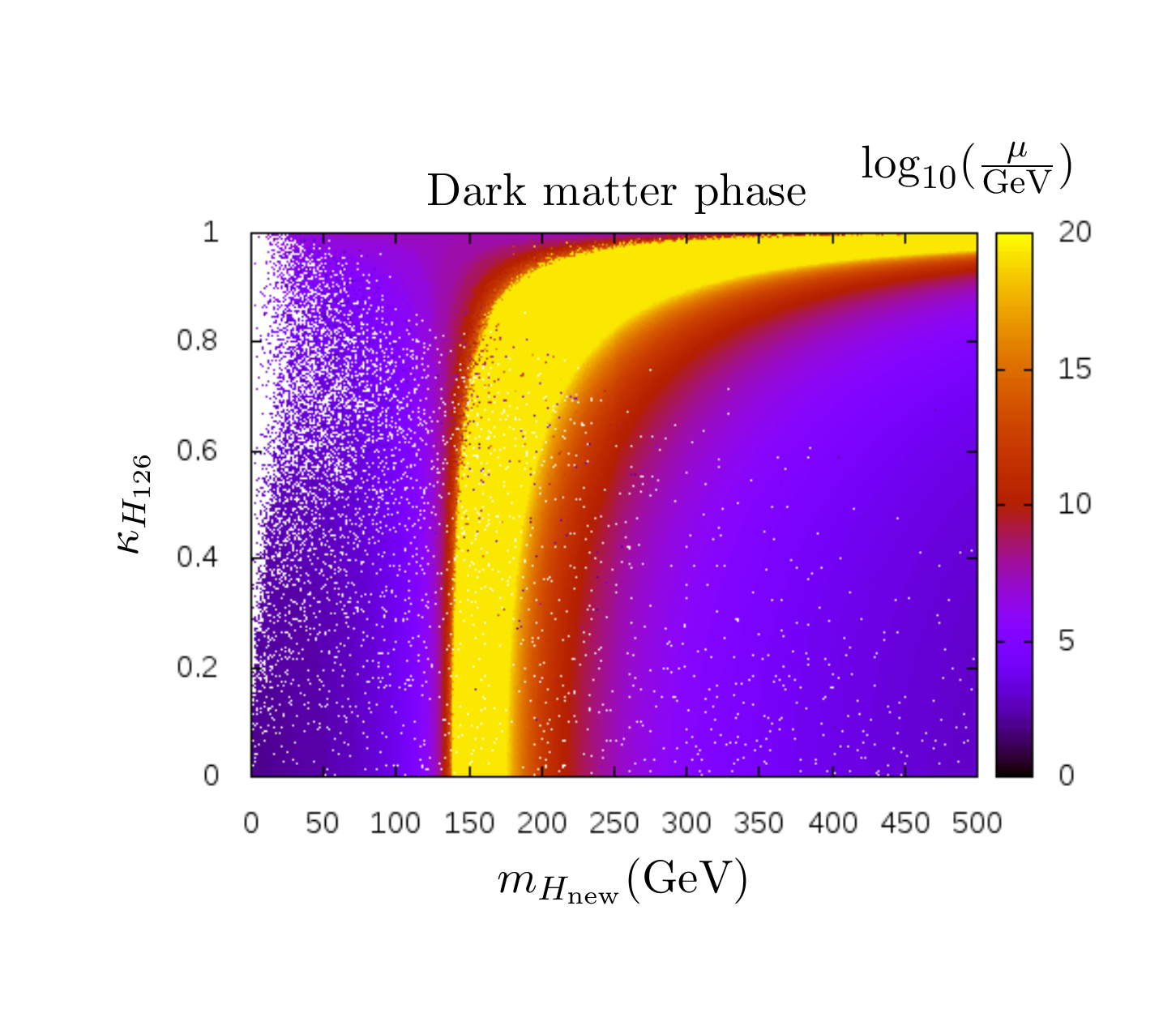} \\
\includegraphics[width=0.35\linewidth,clip=true,trim=40 30 20 50]{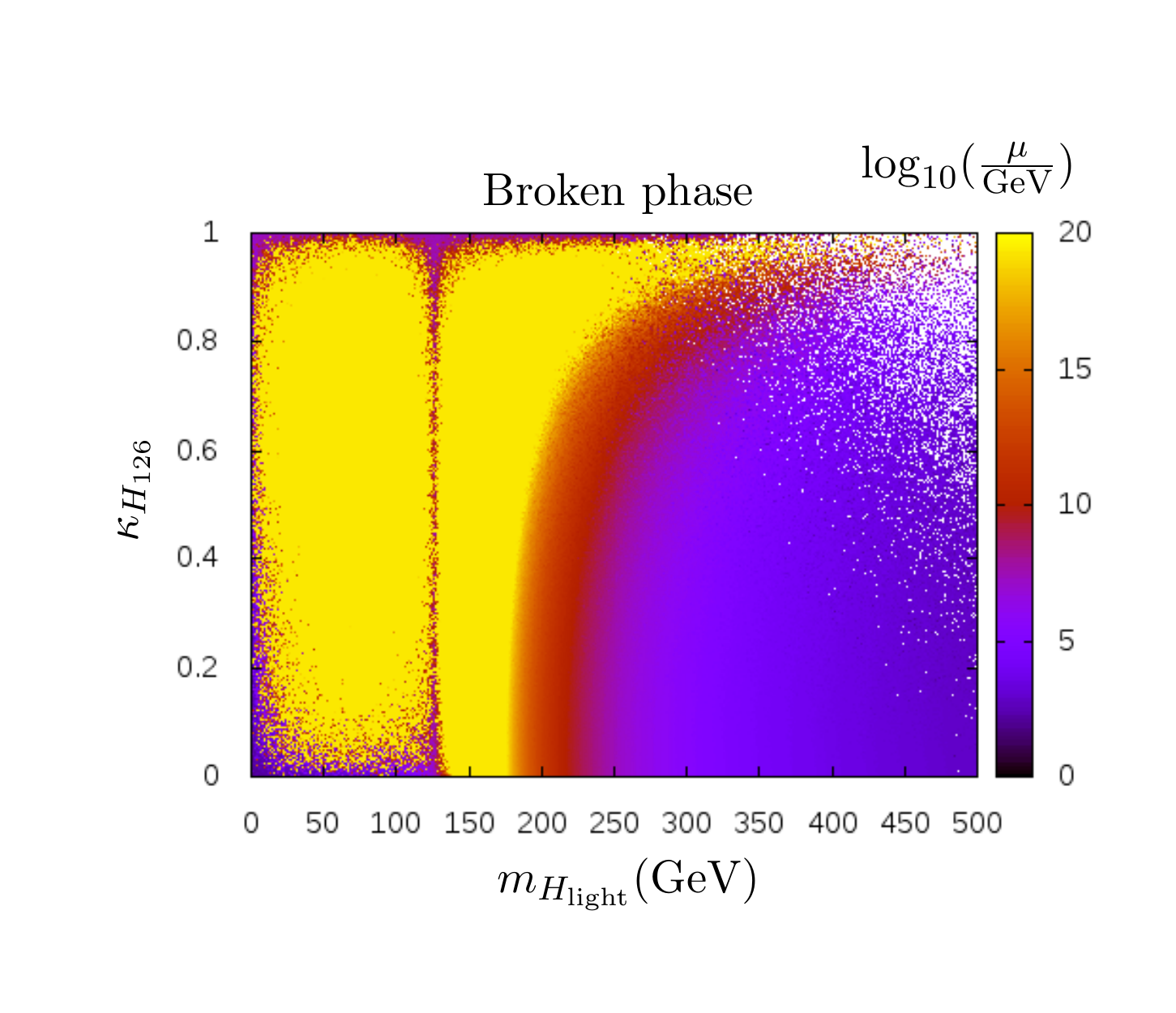}\hspace{0.1\linewidth}\includegraphics[width=0.35\linewidth,clip=true,trim=40 30 20 50]{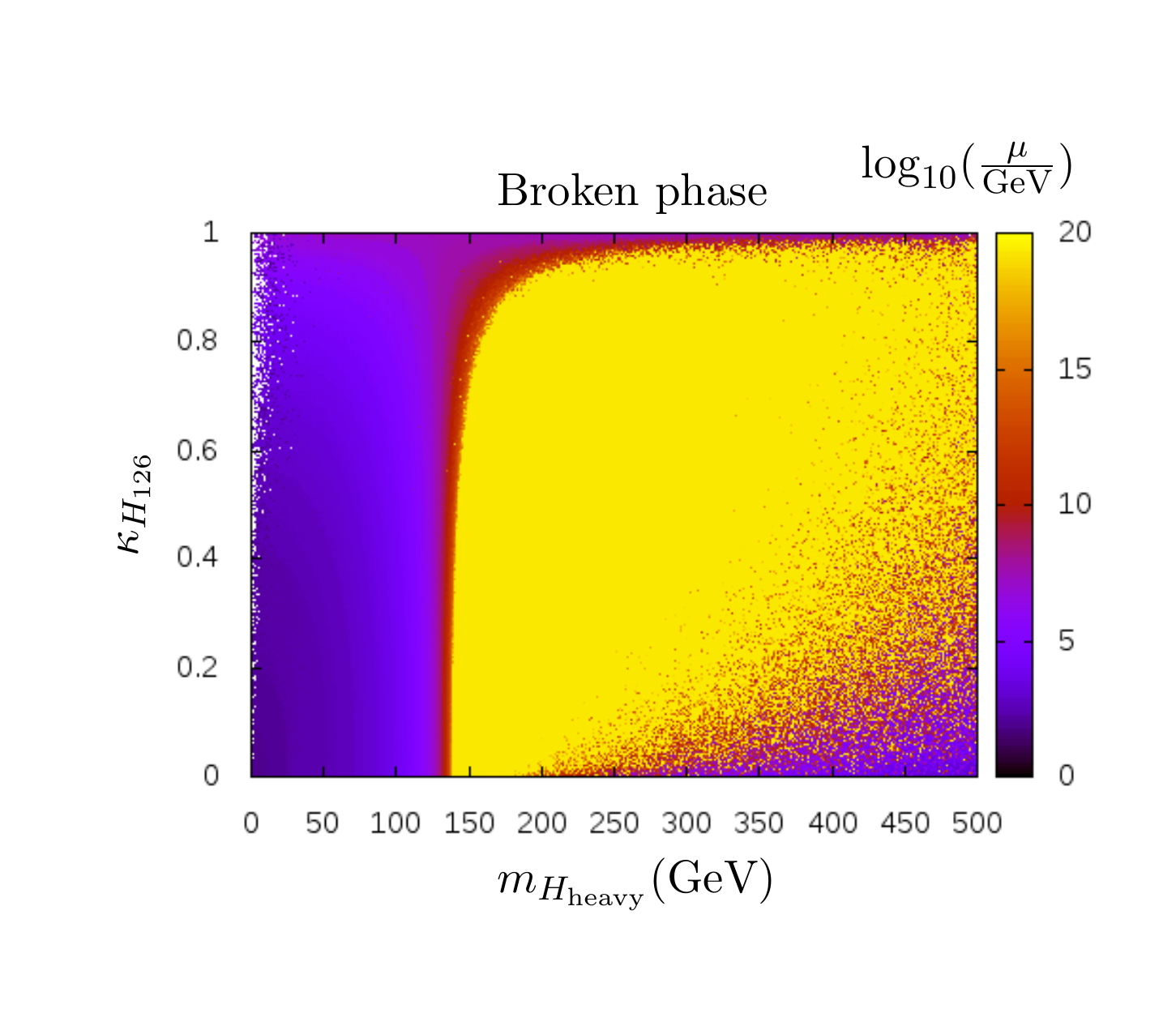}
\end{center}\label{fig:stability_bands}
\caption{Parameter space projections displaying the with the scale up to which the theory remains stable  (in colour) for the dark phase (top panels) and the broken phase (bottom panels). (Adapted from~\cite{Costa:2014qga}).}
\end{figure}
In Fig.~\ref{fig:stability_bands}, we show allowed parameter space points where only theoretical constraints were imposed (vacuum stability, boundedness from below and perturbative unitarity). The colour code of all panels corresponds to the energy scale at which either a runaway direction develops or perturbative unitarity is violated.  On the top left panel, in the vertical axis we have the Higgs quartic coupling versus the mass of the new visible scalar mixing with the Higgs, whereas the right panel contains the coupling suppression of the Higgs found at the LHC. The two boundaries on the left panel are due to the minimum conditions and we can see that there is a yellow band for $0.5<\lambda<1$ for which the model is stable up to the Planck scale which we call the stability band. On the right panel we find that if we impose stability up to an intermediate scale, $10^{10}$ GeV, there is a lower bound on the new mixing scalar mass of roughly 140 GeV.

In the bottom panels we present, for the broken phase, the same coupling versus the mass of the lightest (left) and the heaviest (right)  of the new visible Higgses. The two boundaries of the stability band of the dark phase are now mapped to the upper and lower boundaries of the stability bands of the lightest and the heaviest of the new scalars, respectively. What is interesting to note, is that: i) if a state lighter than 140 GeV is found, the dark matter phase is disfavoured and ii) again, we need one new scalar mixing with the Higgs which is heavier than 140 GeV to take the SM from a metastability scenario into stability.

\begin{figure}
\begin{center}
\includegraphics[width=0.32\linewidth,clip=true,trim=40 40 30 30]{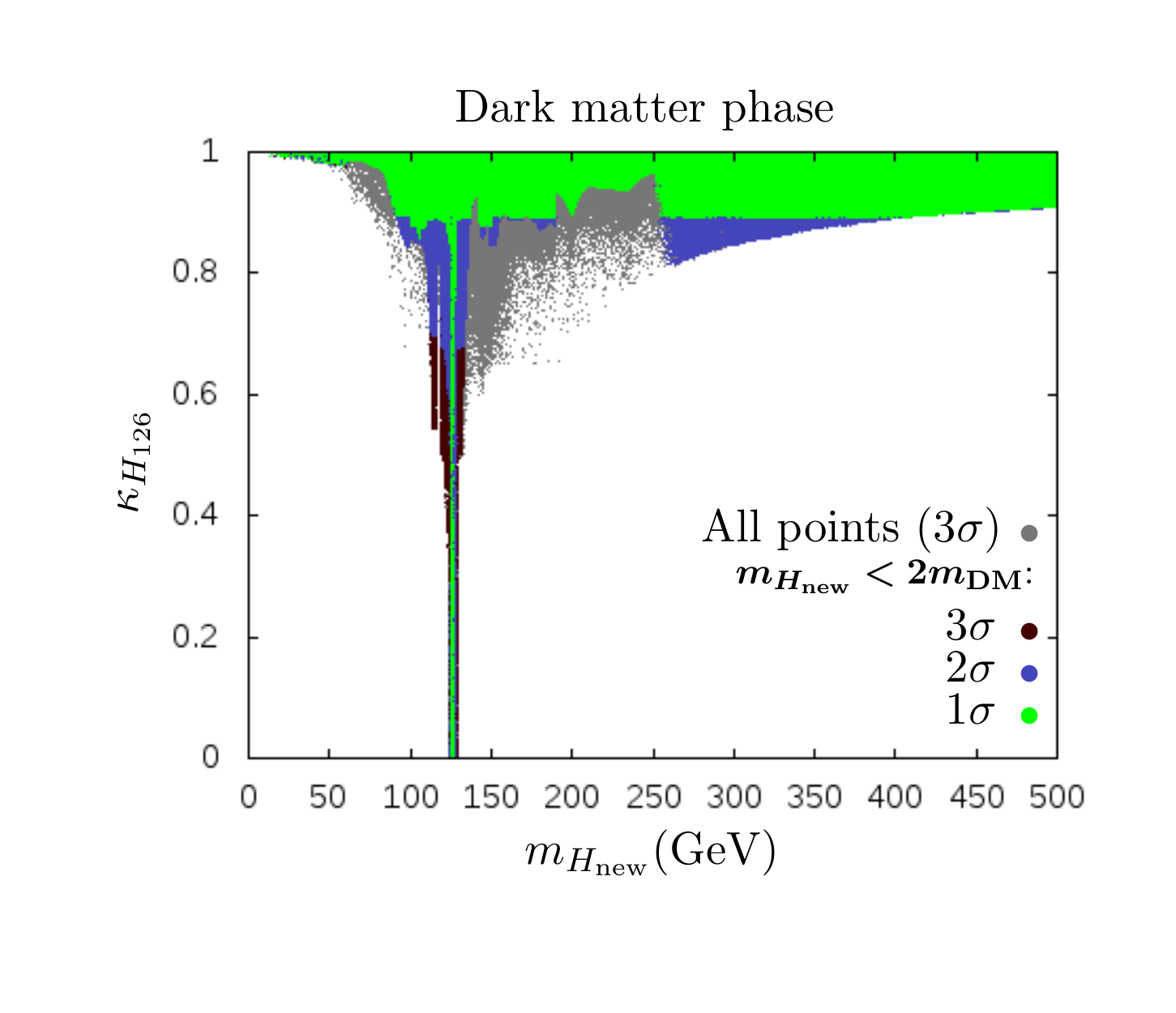} \hspace{0.1\linewidth}
\includegraphics[width=0.37\linewidth,clip=true,trim=30 30 10 30]{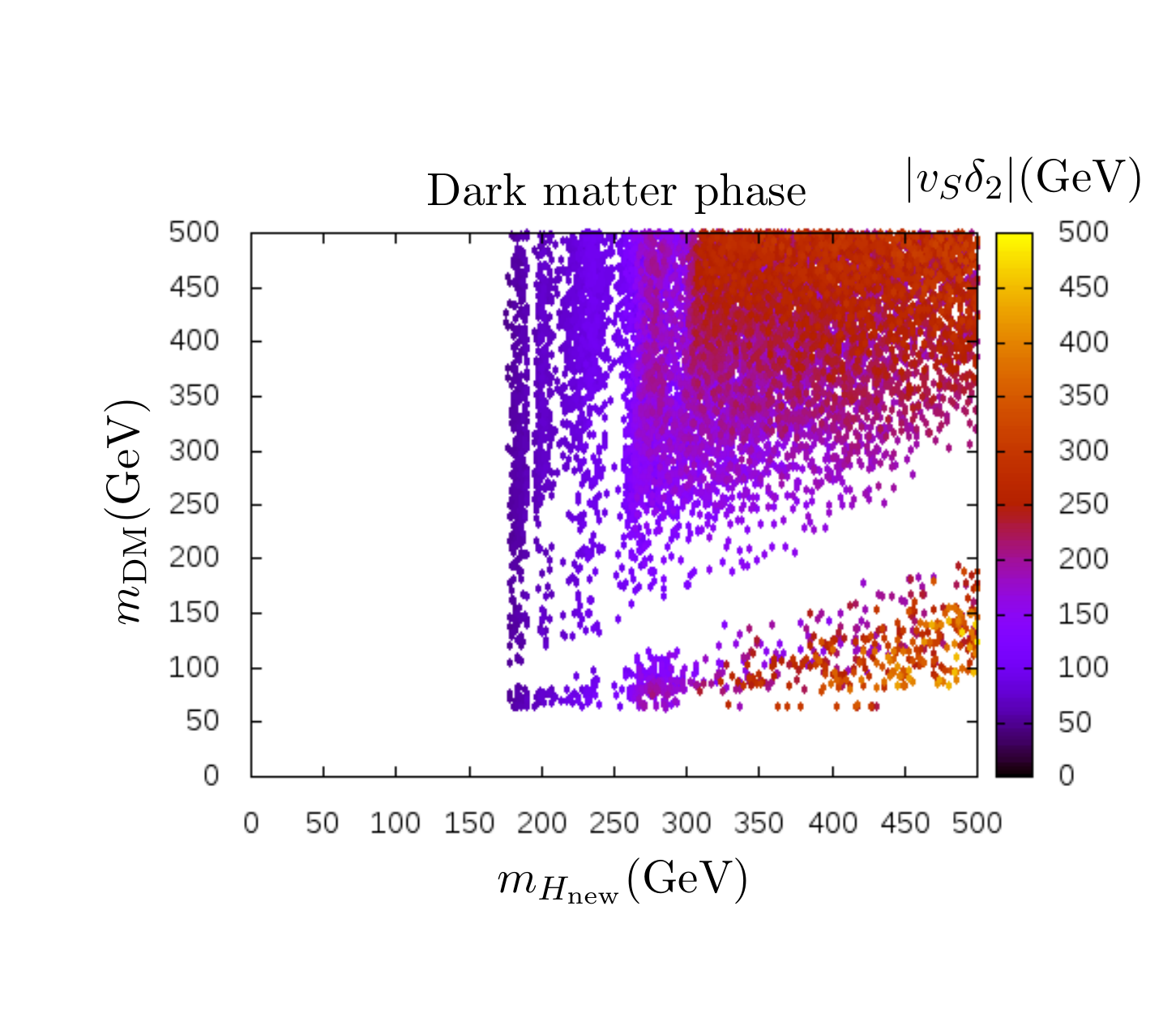}
\end{center}\label{fig_pheno}
\caption{{\em Left:} SM-like Higgs coupling versus  the mass of the new visible particle (in the dark phase) with phenomenological constraints at the weak scale applied, consistent within $1\sigma$, $2\sigma$ and $3\sigma$ of the LHC Higgs measurements. {\em Right:} Combination of all phenomenological constraints with the requirement of stability up to the GUT scale and the correct relic density. (Adapted from~\cite{Costa:2014qga}).}
\end{figure}
Finally, in Fig.~\ref{fig_pheno}, left panel, we present the effect of the phenomenological constraints alone (without RGE running) on the $(\kappa_{126},m_{H_{\rm new}})$ plane for the dark phase. All points are consistent with the Higgs signals at the LHC within $3\sigma$. In the coloured points we have imposed that the new visible scalar cannot decay invisibly to the dark matter candidate (in contrast with the gray points which do not have this cut). In the right panel, we display a projection with the two new scalar masses (dark matter, $m_{\rm DM}$ and $m_{H_{\rm new}}$) where we impose stability up to the GUT scale, and that the model saturates the dark matter relic density measured by Planck (together with the limits from LUX). This pushes the 140 GeV lower bound on the new heavy scalar to roughly 170 GeV\footnote{This limit has been re-discovered recently in a real singlet model in~\cite{Falkowski:2015iwa}} whereas $m_{\rm DM}\gtrsim 50$~GeV.

\section{Conclusions}
We have obtained the two loop RGEs for a complex singlet extension of the SM with dark matter, which may help us improve our understanding of baryogenesis and the stability of the SM at higher energies, and that contains a rich LHC phenomenology (in terms of visible and invisible decays of two new scalar singlet degrees of freedom). In our stability analysis we found that both phases of the model have a stability band where the theory is better behaved at high energies. We also found that, for this to occur, the scalar spectrum must contain a new heavy scalar mixing with the Higgs with mass larger than 140 GeV.

When combined with phenomenology, imposing that the model explains dark matter fully and that it stabilises the theory, the lower bound on the new heavy scalar is pushed up to 170 GeV and the dark particle must be heavier than 50 GeV.

\begin{acknowledgments}
M.S. is grateful to the HPNP2015 organisers for financial support to attend the workshop.  
 The work in this article is supported in part by the Portuguese
\textit{Funda\c{c}\~{a}o para a Ci\^{e}ncia e a Tecnologia} (FCT)
under contract PTDC/FIS/117951/2010. R.S. is also partially supported by PEst-OE/FIS/UI0618/2011. A.M. and M.S. are funded by FCT through the grants SFRH/BPD/97126/2013 and SFRH/BPD/ 69971/2010 respectively. M.S. was also supported by PEst-C/CTM/LA0025/2011.
\end{acknowledgments}

\bigskip 
\bibliography{references}

\end{document}